\documentclass[twocolumn,showpacs,superscriptaddress,preprintnumbers,amsmath,amssymb,nofootinbib]{revtex4}
\usepackage{verbatim}%加注释行用的%
\usepackage{slashed}%输入slash时加载的。 例子：（/slashed{p}）%
\usepackage{epsfig}%加入eps的图形文件%
\usepackage{graphicx}%加入一般图形文件%
\usepackage{amsmath}%数学公式输入宏包%
\usepackage{mathrsfs}%花写字体的输入%
\usepackage{bm}

\begin{document}

\preprint{UB-ECM-PF-10/31}
\preprint{ICCUB-10-053}

%Title of paper
\title{Color-singlet $J/\psi$ production at $\mathcal{O}(\alpha_s^6)$
in $\Upsilon$ decay}

\author{Zhi-Guo He}
\affiliation{Institute of High Energy Physics, Chinese Academy of
Science, P.O. Box 918(4), Beijing, 100049, China.\\
Theoretical Physics Center for Science Facilities,(CAS) Beijing,
100049, China.}

\affiliation{\small{\it{Departament d'Estructura i Constituents de
la Mat\`eria
                   and Institut de Ci\`encies del Cosmos}}\\
        \small{\it{Universitat de Barcelona}}\\
        \small{\it{Diagonal, 647, E-08028 Barcelona, Catalonia, Spain.}}\footnote{Present address}}

\author{Jian-Xiong Wang}

\affiliation{Institute of High Energy Physics, Chinese Academy of
Science, P.O. Box 918(4), Beijing, 100049, China.\\
Theoretical Physics Center for Science Facilities,(CAS) Beijing, 100049, China.}

\date{\today}

\begin{abstract}
% insert abstract here
To clarify the conflict between the theoretical predictions and experimental
measurements of the inclusive $J/\psi$ production in $\Upsilon$ decay,
We consider the $\alpha_s^6$ order color-singlet contributions of
processes  $\Upsilon\to J/\psi+gg$ and $\Upsilon\to J/\psi+gggg$.
Both the branching ratio and $J/\psi$ momentum spectrum are calculated,
and the branching ratio ($4.7\times 10^{-4}$) is larger than the leading-order contribution
($\alpha_s^5$,$\Upsilon\to J/\psi+c\bar{c}g$).
Together with the QCD and QED leading-order contributions
considered in our previous work, the color-singlet prediction of the branching ratio
for the direct $J/\psi$ production is $\mathrm{Br}(\Upsilon\to J/\psi_{\mathrm{direct}}+X)=0.90^{+0.49}_{-0.31}\times10^{-4}$,
which is still about 3.8 times less than the CLEO measurement. We also
obtain a preliminary color-singlet prediction of$R_{cc}=\frac{\mathcal{B}(\Upsilon\to J/\psi+c\bar{c}+X)}
{\mathcal{B}(\Upsilon\to J/\psi+X)}$ and find the value $0.39^{+0.21}_{-0.20}$ is much
larger than the color-octet predictions, and suggest to measure this quality in future experimental analysis.

\end{abstract}

% insert suggested PACS numbers in braces on next line
%\begin{keyword}
%\end{keyword}
% insert suggested keywords - APS authors don't need to do this
%\keywords{}
%\maketitle must follow title, authors, abstract, \pacs, and \keywords
\maketitle

% body of paper here - Use proper section commands
% References should be done using the \cite, \ref, and \label commands

The existence of a hierarchy of energy scales: $m_Q>>m_Q v>>m_Q v^2$ makes the
heavy quarkonium system be an ideal laboratory to study both perturbative and
nonperturbative aspects of QCD, where $v$, being assumed to be much smaller than
$1$, is the velocity of heavy quark in the rest frame of the heavy meson.
And it is commonly believed that the nonrelativistic QCD (NRQCD)\cite{Bodwin:1994jh}
effective theory provides a rigorous factorization formulism to separate the
physics in different scales. The NRQCD not only covers the results of previous
color-singlet(CS) model predictions, where at short distance the $Q\bar{Q}$ can
only be in the color-singlet configuration with the same quantum numbers as the
corresponding heavy quarkonium states, it also includes the contribution of
$Q\bar{Q}$ in the color-octet(CO) configuration at short distance.

Despite of the impressive success of the NRQCD, the role of the CO is not well
established yet, particularly in $J/\psi$ production case. The substantial theoretical
progress in the next-to-leading-order(NLO) calculations shows that there is not
a convincing mechanism to explain the $J/\psi$ production data in varies experiments
self-consistently yet. For $J/\psi+c\bar{c}+X$\cite{Abe:2002rb} and
$J/\psi+X_{\mathrm{non-c\bar{c}}}$\cite{:2009nj} production in $e^{+}e^{-}$ annihilation
at B-factories, the CS processes themselves can account for the cross sections when the
NLO QCD corrections\cite{Zhang:2005cha,Gong:2009ng,Ma:2008gq,Gong:2009kp} and relativistic
corrections \cite{He:2007te,He:2009uf} are taken into account, which leaves very little
room for the CO contribution\cite{Zhang:2009ym}.
For $J/\psi$ production from $Z$ decay, recent NLO QCD correction calculation in the CS
\cite{Li:2010xu} gives just one-half of the experimental measurement and the CO contribution
might be able to explain the other-half. The transverse momentum $p_t$ distribution
of $J/\psi$ photoproduction and polarization parameters at HERA can not be well described by
the CS\cite{Artoisenet:2009xh,Chang:2009uj} at QCD NLO, it seems that $p_t$ distribution
of $J/\psi$ photoproduction can be explained by the CO and CS contribution together
at QCD NLO\cite{Butenschoen:2009zy};
For $J/\psi$ hadroproduction, together with the CS \cite{Campbell:2007ws,Gong:2008sn,Gong:2008hk}
contribution and the CO contribution \cite{Braaten:1994vv,Gong:2008ft} at NLO in $\alpha_s$, we
can not describe the Tevatron results about the $p_t$ distribution of $J/\psi$ production and polarization
simultaneously yet.

In order to clarify such a puzzling theoretical situation, it is worth to further investigate
some other $J/\psi$ production processes, one of which is the inclusive $J/\psi$ production
in $\Upsilon$ decay. From the theoretical point of view, because $\Upsilon$ predominately decays
into three gluons via $b\bar{b}$ annihilation, it is proposed\cite{Cheung:1996mh,Napsuciale:1997bz}
that in the rich-gluon final state environment abundant $J/\psi$ can be produced through
$c\bar{c}$ pair in CO $^3S_1$ configuration. Hence, the inclusive $J/\psi$ production in
$\Upsilon$ decay will be an another good probe to discriminate the CS and CO mechanism.
And the present CO predictions of the branching ratio is
$\mathcal{B}(\Upsilon\to J/\psi+X)=6.2\times10^{-4}$ \cite{Cheung:1996mh,Napsuciale:1997bz}
with about $10\%$ feed-down contribution from $\psi(2S)$ and $10\%$ from $\chi_{cJ}$
\cite{Trottier:1993ze}. While the correct CS result, which was overestimated by about an order
in magnitude before \cite{Li:1999ar,Han:2006vi}, is only $4.2\times10^{-5}$\cite{He:2009by}.
On the experimental side, the branching ratio for $\Upsilon\to J/\psi+X$
has been measured by a few collaborations about twenty years ago
\cite{Fulton:1988ug,Maschmann:1989ai,Albrecht:1992ap}, and recently, the more precise
measurement carried by the CLEO Collaboration gave\cite{Briere:2004ug}
\begin{equation}\label{exp}
\mathcal{B}(\Upsilon\to J/\psi+X)=(6.4\pm0.4\pm0.6) \times10^{-4}.
\end{equation}

It can be seen that the CLEO result is in good
agreement with the CO prediction, but the $J/\psi$ momentum distribution
measured by CLEO \cite{Briere:2004ug} is much softer than the CO predictions
\cite{Cheung:1996mh,Napsuciale:1997bz}. In a very recent work\cite{Liu:2009ra}, it
is found that the momentum spectrum can be significantly softened after combining
the NRQCD and Soft Collinear Effective Theory(SECT) in the kinematic endpoint region.
However, it yields a much smaller branching ratio. This may indicate that the CO
processes do not contribute dominantly.

In the CS case, the results at QCD and QED leading-order(LO)\cite{He:2009by} is about
an order of magnitude smaller than the experimental result.  To investigate
the CS contributions for the $J/\psi$ production in $\Upsilon$ decay more precisely,
we consider the $\mathcal{O}(\alpha_s^6)$ contributions through $\Upsilon\to J/\psi+gg$
and $\Upsilon\to J/\psi+gggg$ processes.
It was refereed in Ref.\cite{Trottier:1993ze} that this two processes had been crudely
estimated and the branching ratio is a few of $\times$ $10^{-4}$, which is comparable to
CLEO data. Since the former calculation refereed is rough without any publication and detail,
it is necessary to give an exact result and a complete analysis of these two processes.
To perform the calculations, we employ the Feynman Diagram Calculation (FDC) package~\cite{FDC}.

According to NRQCD factorization approach, at the leading order of $v_b^2$ and $v_c^2$,
the CS contribution to $\Upsilon\to J/\psi+X$ is expressed as:
\begin{eqnarray}\label{fac}
&&d\Gamma(\Upsilon\to
J/\psi+X)=d\hat{\Gamma}(b\bar{b}[^3S_1,\underline{1}]\to\nonumber\\&&
c\bar{c}[^3S_1,\underline{1}]+X)
\times\langle\Upsilon|\mathcal{O}_1(^3S_1)|\Upsilon\rangle
\langle\mathcal{O}^{\psi}_1(^3S_1)\rangle,
\end{eqnarray}
where $d\hat{\Gamma}$ is partonic partial decay width which can be calculated
perturbatively. By dimension analysis, it is easy to derive out that the general
expression of the partial width is written as
\begin{equation}\label{fac1}
\hat{\Gamma}=\frac{1}{3(2N_c)^2}\frac{\alpha_s^6}{m_b^{5}}f(r)
\end{equation}
where $r=m_c/m_b$ is a dimensionless parameter and $f$ is a process dependent
function of $r$. The $\langle\Upsilon|\mathcal{O}_1(^3S_1)|\Upsilon\rangle$
and $\langle\mathcal{O}^{\psi}_1(^3S_1)\rangle$ in Eq.[\ref{fac}] are the nonperturbative
matrix elements, which will be determined phenomenologically. To be consistent with
our former work \cite{He:2009by}, we keep the factor $\frac{1}{3(2N_c)^2}$ explicitly.

\begin{figure}
\begin{center}
\includegraphics[scale=0.45]{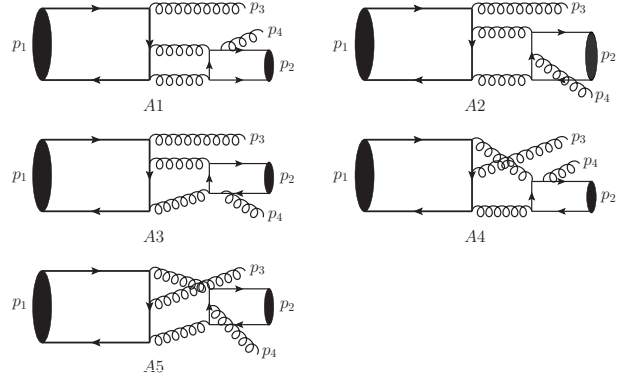}
\caption{The typical Feynman diagrams of five of ten groups in the CS $\Upsilon\to J/\psi+gg$ process,
and the other five typical ones can be obtained by exchanging the positions of the two final-state gluons.}
\end{center}
\end{figure}

The $\mathcal{O}(\alpha_s^6)$ $\Upsilon\to J/\psi+gg$ process is of
36 one-loop Feynman Diagrams at leading-order. To simplify the calculation,
each diagram is summed up with it's possible
partner diagram which is obtained by reversing the direction of b-quark or c-quark line in fermion loop.
Then the diagrams are divided into ten groups, and the representative ones
are shown in Fig.[1] and others can be obtained by exchanging the positions of the two final-state gluons.
For diagrams in the same group, the amplitude equals to each other when ignoring color factor,
thus only the $d^{abc}d^{abc}$ piece in color factor will survive after all the diagrams
being summed up in each group. It is found that
there are infrared divergences in the amplitudes of diagrams in A1, A2, A4, A5 groups, at least
in Feynman gauge. And the divergence terms in A1(A2) group cancel those in A4(A5), then
the total amplitude is finite. The amplitude of each diagram in group A3 is
finite individually because in such diagrams there are no such a virtual gluon which joints
two on shell (anti)quarks or one on shell quark and one on shell antiquark.

Before showing the result, we would like to address some non-trivial techniques treatment in the
calculation. By applying the FDC package, the general expression of the Feynman amplitude for
$\Upsilon(p_1,\epsilon_1)\to J/\psi(p_2,\epsilon_2) + g(p_3,\epsilon_3) + g(p_4,\epsilon_4)$
process is generated as:
\begin{widetext}
\begin{equation}\label{GE}
\begin{array}{ll}
 M=&\epsilon_{1}\cdot \epsilon_{2}\epsilon_{3}\cdot
\epsilon_{4}c_{41}+\epsilon_{1}\cdot \epsilon_{3}
\epsilon_{2}\cdot \epsilon_{4}c_{42}+\epsilon_{1}\cdot
\epsilon_{4}\epsilon_{2}\cdot \epsilon_{3}c_{43}+{p_4}\cdot
\epsilon_{2}{p_4}\cdot \epsilon_{3}\epsilon_{1}\cdot
\epsilon_{4}c_{19}+{p_4}\cdot \epsilon_{1}(  {p_4}
\cdot \epsilon_{2}\epsilon_{3}\cdot \epsilon_{4}c_{37}
\\&+{p_4} \cdot \epsilon_{3}\epsilon_{2}\cdot \epsilon_{4}c_{39})+
{p_3}\cdot \epsilon_{4}[{p_4}\cdot \epsilon_{1}(\epsilon_{2}
\cdot \epsilon_{3}c_{21}+{p_4}\cdot \epsilon_{2}{p_4}\cdot
\epsilon_{3}c_{9})+{p_4}\cdot \epsilon_{2}\epsilon_{1}\cdot
\epsilon_{3}c_{29} +{p_4}\cdot \epsilon_{3}
\epsilon_{1}\cdot \epsilon_{2}c_{31}]
\\&+{p_3}\cdot
\epsilon_{2}[{p_4}\cdot \epsilon_{3}\epsilon_{1}\cdot
\epsilon_{4}c_{20}+{p_4}\cdot \epsilon_{1}\epsilon_{3}\cdot
\epsilon_{4}c_{35}+{p_3}\cdot \epsilon_{4}(\epsilon_{1}
\cdot \epsilon_{3}c_{27}+{p_4}\cdot \epsilon_{1}{p_4}\cdot
\epsilon_{3}c_{10})]+{p_3}\cdot \epsilon_{1}[  {p_4}
\cdot \epsilon_{2}\epsilon_{3}\cdot \epsilon_{4}c_{38}
\\& +{p_4} \cdot \epsilon_{3} \epsilon_{2}\cdot \epsilon_{4}c_{40}+{p_3}
\cdot \epsilon_{4}(\epsilon_{2}\cdot \epsilon_{3}c_{22}+
{p_4}\cdot \epsilon_{2}{p_4}\cdot \epsilon_{3}c_{11})+{p_3}
\cdot \epsilon_{2}(\epsilon_{3}\cdot \epsilon_{4}c_{36}
+{p_3}\cdot \epsilon_{4}{p_4}\cdot \epsilon_{3}c_{12 })]
\\& +{p_2}\cdot \epsilon_{3}\{{p_4}\cdot \epsilon_{2}
\epsilon_{1}\cdot \epsilon_{4}c_{17}+{p_4}\cdot \epsilon_{1}
\epsilon_{2}\cdot \epsilon_{4}c_{33}+{p_2}\cdot \epsilon_{4}
[\epsilon_{1}\cdot \epsilon_{2}c_{26}+{p_3}\cdot
\epsilon_{2}{p_4}\cdot \epsilon_{1}c_{6}+{p_3}\cdot
\epsilon_{1}({p_3}\cdot \epsilon_{2}c_{8}
\\& +{p_4}\cdot \epsilon_{2}c_{7}) +{p_4}\cdot \epsilon_{1}{p_4}\cdot
\epsilon_{2}c_{5}]+{p_3}\cdot \epsilon_{4}(\epsilon_{1}
\cdot \epsilon_{2}c_{25}+{p_4}\cdot \epsilon_{1}{p_4}\cdot
\epsilon_{2}c_{1}) +{p_3}\cdot \epsilon_{2}(
\epsilon_{1}\cdot \epsilon_{4}c_{18}+{p_3}\cdot \epsilon_{4}
{p_4}\cdot \epsilon_{1}c_{2})
\\& +{p_3}\cdot \epsilon_{1}
(\epsilon_{2}\cdot \epsilon_{4}c_{34} +{p_3}\cdot \epsilon_{4}
{p_4}\cdot \epsilon_{2}c_{3}+{p_3}\cdot \epsilon_{2}{p_3}
\cdot \epsilon_{4}c_{4})\}+{p_2}\cdot \epsilon_{4}[
{p_3}\cdot \epsilon_{2}(\epsilon_{1}\cdot \epsilon_{3}c_{28}
+{p_4}\cdot \epsilon_{1}{p_4}\cdot \epsilon_{3}c_{14})
\\& +{p_3} \cdot \epsilon_{1}(\epsilon_{2}\cdot \epsilon_{3}c_{24}
+{p_3}\cdot \epsilon_{2}{p_4}\cdot \epsilon_{3}c_{16}+{p_4}
\cdot \epsilon_{2}{p_4}\cdot \epsilon_{3}c_{15}) +
{p_4}\cdot \epsilon_{1}(\epsilon_{2}\cdot \epsilon_{3}c_{23}
+{p_4}\cdot \epsilon_{2}{p_4}\cdot \epsilon_{3}c_{13})
\\& +{p_4}
\cdot \epsilon_{2}\epsilon_{1}\cdot \epsilon_{3}c_{30}+{p_4}
\cdot \epsilon_{3}\epsilon_{1}\cdot \epsilon_{2}c_{32}]
\end{array}\end{equation}
\end{widetext}
where $c_i,~i=1,\dots,43$ are the coefficients of the Lorentz structure and all the loop diagrams
contribute to their values. From the general expression in Eq.(\ref{GE}), it is easy to understand
why the tensor reduction procedures are very complicated and will generate complicated results for
the coefficients $c_i$.

These complicated results of $c_i$ may contain fake pole structures and
cause big number cancellation problem, finally will spoil numerical calculation in limited precision case.
In fact, it really took place in our numerical calculation. We found it is impossible to control
the cancellation of the big numbers and to obtain correct results in double precision FORTRAN calculation.
In quadruple precision FORTRAN calculation, we can obtain the correct results by introducing cut-conditions
to control these fake poles in phase space integration. To demonstrate that our treatment is suitable,
we define a cut condition parameter in phase space integration as
\begin{equation}
A = \frac{\Gamma(|M|^2=1,~with~cuts)}{\Gamma(|M|^2=1,~without~cuts)}
\end{equation}
and calculate the partial decay width numerically with different cut condition parameter.
The results are:
\begin{equation}
\begin{array}{ll}
&\Gamma=a(1.440\pm0.001)\times 10^{-10}~for~A=1-0.04 \\
&\Gamma=a(1.446\pm0.001)\times 10^{-10}~for~A=1-0.005 \\
&\Gamma=a(1.450\pm0.001)\times 10^{-10}~for~A=1-0.0005,
\end{array}
\end{equation}
where all the calculations are under control in quadruple precision FORTRAN calculation,
and $a$ is just a constant number.
We have tried to do the calculation with $A=1-0.00005$ and found a divergent result.
It means that the big number cancellation will lose control even in quadruple precision FORTRAN
when $A$ is too close to $1$. Therefore, our following calculations for $\Upsilon\to J/\psi+gg$
are based on the cut condition parameter $A=1-0.0005$.
Moreover, to check the gauge invariance, in the expression of Eq.(\ref{GE}) we replace the
gluon polarization vector $\epsilon_3$ (or $\epsilon_4$) by its 4-momentum $p_3$ (or $p_4$)
in the final numerical calculation.  Definitely the result must be zero and
our results reproduce it.

Now, we proceed to present our results. Since the two interior gluons
can be on shell simultaneously in this process, the amplitude will be complex-valued. And we use the
superscript `Im" to denote the contribution of the real process $\Upsilon\to 3g$ followed
by $gg\to J/\psi+g$. Setting $m_c=m_{J/\psi}/2=1.548$ GeV and $m_b=m_{\Upsilon}/2=4.73$GeV,
which corresponds to $r=1.548/4.73=0.327$, we get $f_{gg}(r)=2.07$ and $f_{gg}^{\mathrm{Im}}(r)=0.741$
which is about $1/3$ of the total. To show the dependence of $f$ on $r$, we
also list some of the numerical results of $f(r)$ in Tab.I, where $r$ is in the range of
$0.275<r<0.381$, which is obtained by fixing the value of $m_b$ and varying $m_c$ from 1.3GeV to
1.8GeV GeV\cite{Li:1999ar}. For comparison we also list the results of $f_{ccg}(r)$ for the
$\mathcal{O}(\alpha_s^5)$ $\Upsilon\to J/\psi+c\bar{c}+g$ process. It can be seen that both the
value of $f_{gg}(r)$ and that of $f_{gg}^{\mathrm{Im}}(r)$ do not change sharply, when $r$
goes from $0.275$ to $0.381$, and this behavior is quite different from what happens to $f_{ccg}(r)$.
Also the ratio of $f_{gg}^{\mathrm{Im}}(r)$ to $f_{gg}(r)$ changes very little with $r$.

\begin{table}
\caption{The values of $f(r)$ for $J/\psi+c\bar{c}+g$ ($f_{ccg}$), $J/\psi+gg$
($f_{gg}$ and $f_{gg}^{\mathrm{Im}}$),$J/\psi+gggg$ ($f_{4g}$) production
in $\Upsilon$ decay with different inputs of $r=\frac{m_c}{m_b}$.}
\begin{tabular}{|c|c|c|c|c|}
\hline    r  & $f_{ccg}(r)$&$f_{gg}(r)$& $f_{gg}^{\mathrm{Im}}(r)$& $f_{4g}(r)$\\
\hline~ 0.275&0.904&2.94&1.02&~$1.35\times10^{-2}$~\\
\hline~ 0.296&0.567&2.54&0.892&$1.08\times10^{-2}$\\
\hline~ 0.317&0.345&2.21&0.786&$0.880\times10^{-2}$\\
\hline~ 0.327&0.269&2.07&0.741&$0.800\times10^{-2}$\\
\hline~ 0.338&0.202&1.94&0.696&$0.721\times10^{-2}$\\
\hline~ 0.361&0.105&1.68&0.612&$0.585\times10^{-2}$\\
\hline~ 0.381&0.055&1.49&0.547&$0.490\times10^{-2}$\\
\hline
\end{tabular}
\end{table}

There are 216 Feynman diagrams in the CS $\Upsilon\to J/\psi+gggg$ process, and the typical one is
shown in Fig.2. It is a tree process without infrared divergence, and the numerical
results are calculated straightforwardly with the help of FDC package. When $r=0.327$, we get
$f_{4g}(r)=0.8\times10^{-2}$, which is more than two orders less than $f_{gg}(0.327)$. Some other
numerical results of $f_{4g}(r)$ for $0.275<r<0.381$ are also listed in Tab.I. Like $f_{gg}(r)$,
the function $f_{4g}(r)$ also does not dependent on r seriously, but its value is too small
comparing to the values of the $f$ functions of the other processes. One possible reason
is that the five-body phase space is much smaller than the three-body phase space.

Besides $f_{gg}(r)$($f_{4g}(r))$, the partial decay width $\Gamma(\Upsilon\to J/\psi+gg(4g))$ also
dependents on the choice of the values of the two NRQCD long-distance matrix elements, the coupling
constant $\alpha_s$ and the b-quark mass $m_b$. The value of
$\langle\mathcal{O}_1^{\psi}(^3S_1)\rangle\simeq 3\langle J/\psi|\mathcal{O}(^3S_1)|J/\psi\rangle$
can be extracted from $J/\psi$ decay into $e^{+}e^{-}$ by using the upto $\alpha_s$ order result
\begin{equation}\label{matrix1}
\Gamma(J/\psi\to e^{+}e^{-})=\frac{ 2\pi e_{c}^2\alpha^2}{3m_c^2}(1-\frac{16\alpha_s}{3\pi})
\langle\psi|\mathcal{O}_1(^3S_1)|\psi\rangle.
\end{equation}
 Using $\alpha=1/128$, $m_c=1.548\mathrm{GeV}$, $\alpha_s(2m_c)=0.26$,
$\Gamma(J/\psi\to e^{+}e^{-})=5.54\mathrm{keV}$\cite{Amsler:2008zzb},
we get $\langle\mathcal{O}_1^{\psi}(^3S_1)\rangle=1.25\mathrm{GeV}^{3}$. And
$\langle\Upsilon|\mathcal{O}(^3S_1)|\Upsilon\rangle=2.92\mathrm{GeV}$ is be determined in a similar
way with  $m_b=4.73\mathrm{GeV}$, $\alpha_s(2m_b)=0.18$,
$\Gamma(\Upsilon\to e^{+}e^{-})=1.29\mathrm{keV}$\cite{Amsler:2008zzb}.
The uncertainty from the choice of the renormalization scale is quite large since there are two typical energy
scales $m_b$ and $m_c$ in the calculation. By choosing $\mu=2m_c$,
we find
\begin{eqnarray}
\Gamma^{g}&=\Gamma(\Upsilon\to J/\psi+gg)+\Gamma(\Upsilon\to J/\psi+4g)\nonumber\\
&=9.1\times10^{-3}\mathrm{keV}
\end{eqnarray}
It corresponds to
\begin{eqnarray}
\mathcal{B}^{g}&=\mathcal{B}(\Upsilon\to J/\psi+gg)+\mathcal{B}(\Upsilon\to J/\psi+4g)\nonumber\\
               &=1.7\times10^{-4}
\end{eqnarray}
which is coincident with the rough result mentioned in Ref.\cite{Trottier:1993ze}. However,
the branching ratio becomes much smaller and is only $2.32\times10^{-5}$ when choosing $\mu=2m_b$.
This is because the processes are at $\alpha_s^6$ order. Note that, seemingly, the branching ratio
also strongly dependent on $m_b$ as $m_b^{-5}$ , but in fact the dependence
is $m_b^{-3}$ because of the  dependence of the nonperturbative matrix element
$\langle\Upsilon|\mathcal{O}(^3S_1)|\Upsilon\rangle$ on $m_b$ from it's phenomenological determination.
To obtain the above numerical results for the branching ratio, the experimental measurement on the $\Upsilon$
total decay width $\Gamma_{\Upsilon}=53\mathrm{keV}$ \cite{Amsler:2008zzb} is used.

\begin{figure}
\begin{center}
\includegraphics[scale=0.80]{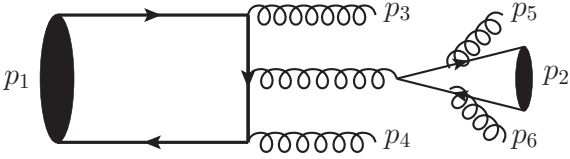}
\caption{One of the 216 Feynman diagrams for the CS $\Upsilon\to J/\psi+gggg$
process.}
\end{center}
\end{figure}

Since the $J/\psi$ is produced through the three-gluon decay channel of $\Upsilon$ in these two processes,
It is natural to normalize the partial width to the decay width of $\Upsilon\rightarrow ggg$, which at LO in $\alpha_s$
is given by
\begin{equation}\label{bdecay}
\Gamma(\Upsilon\to ggg)=\frac{20\alpha_s^3}{243m_b^2}(\pi^2-9)
\langle\Upsilon|\mathcal{O}(^3S_1)|\Upsilon\rangle.
\end{equation}
Then branching ratio is expressed in an alternate form
\begin{equation}\label{ratio}
\mathcal{B}^{g}=\Gamma_{\mathrm{Nor}}^{g}\times \mathcal{B}(\Upsilon\to ggg),
\end{equation}
where
\begin{equation}\label{normal}
\Gamma_{\mathrm{Nor}}^{g}=\frac{81(f_{gg}(r)+f_{4g}(r))\alpha_s^3\langle\mathcal{O}_1^{\psi}(^3S_1)\rangle}
{20(2N_c)^2m_b^3(\pi^2-9)},
\end{equation}
and $\mathcal{B}(\Upsilon\to ggg)=84\%$ is obtained by assuming
$\mathcal{B}(\Upsilon\to ggg)\approx\mathcal{B}(\Upsilon\to \mathrm{light\; hadron\;(LH)})$
\footnote{The contribution of $\Upsilon\to\gamma^{\ast}\to q\bar{q}$ is excluded.}\cite{Amsler:2008zzb}.
To calculate the branching ratio in this way is equivalent to determine
$\alpha_s^3\langle\Upsilon|\mathcal{O}(^3S_1)|\Upsilon\rangle$ from LH decay of $\Upsilon$ and can reduce the
uncertainties from $\alpha_s$. And our following results are all calculated based on Eq.[\ref{ratio}].

The numerical results of ``f(r)" for each decay process are presented in Tab.1 and it show that the values of
$f_{(gg)}(r)$ changes slowly when r goes from 0.275 to 0.381.
The Feynman diagrams in Fig.[1] indicate the process $\Upsilon\to J/\psi+gg$
can be viewed as $\Upsilon\to gg^{(\ast)}g^{(\ast)}$\footnote{$g^{(\ast)}$ means the gluon can either be virtual
or real.} followed by $g^{(\ast)}g^{(\ast)}\to J/\psi+g$. Then the normalized $\Gamma_{\mathrm{Nor}}^{g}$can
cancel part of the contribution at $m_b$ scale, so similar to what is done in Ref.\cite{Cheung:1996mh}
we choose the scale of $\alpha_s$ to be $2m_c$. The theoretical uncertainties of $\Upsilon\to J/\psi+gggg$
can be analyzed in the same way.

By setting the default parameter choice: $m_b=4.73\mathrm{GeV}$, $r=0.327$,
$\langle\mathcal{O}_1^{\psi}(^3S_1)\rangle=1.25\mathrm{GeV}^{3}$ and
$\alpha_s(2m_c)=0.26$, we obtain
\begin{equation}\label{bran1}
\mathcal{B}^{g}=\Gamma_{\mathrm{Nor}}^{g}\times \mathcal{B}(\Upsilon\to ggg)=0.47\times10^{-4}.
\end{equation}
Using the same inputs to re-estimated the result in Ref.\cite{He:2009by} and adding up it with
the contribution in Eq.[\ref{bran1}], we obtain the total CS singlet prediction
\begin{equation}
\mathcal{B}(\Upsilon\to J/\psi+X)=0.90\times10^{-4},
\end{equation}
where the total contribution from the $\mathcal{O}(\alpha_s^6)$ $J/\psi+gg$ and $J/\psi+gggg$  processes is
as important as those calculated in Ref.\cite{He:2009by}.
It is clear that the uncertainties are from the b quark mass $m_b$, the scaleless functions
$f_{gg}(r)$ and $f_{4g}(r)$, and the choice of the scale of $\alpha_s$.
To estimate the uncertainty, we used $m_b=4.6GeV,~r=0.296$ and $\mu=2m_c$ for upper bound;
$m_b=4.9GeV,~r=0.361$ and $\mu=2m_c$ for lower bound, then the branching ratio is represented as:
%{\bf (ZG, I calculated the uncertainty from $[f_{c\bar{c}g}(r)/\alpha_s + f_{gg}(r)+f_{4g}(r)]$ part,
%Please add the  remaining part uncertainty to the following number)}
\begin{equation}
\mathcal{B}(\Upsilon\to J/\psi+X)=0.90^{+0.49}_{-0.31}\times10^{-4}.
\end{equation}
Furthermore the total branching ratio turns to be a much smaller value $6.3\times10^{-5}$ $(5.2\times10^{-5})$
by choosing the scale to be $2m_b$ ($2\sqrt{m_bm_c}$) and $\alpha_s(2m_b)=0.18$ ($\alpha_s(2\sqrt{m_cm_b})=0.21$)
and keeping the other parameters the same as for the central value of the branching ratio.
\begin{figure}
\begin{center}
\includegraphics[scale=0.80]{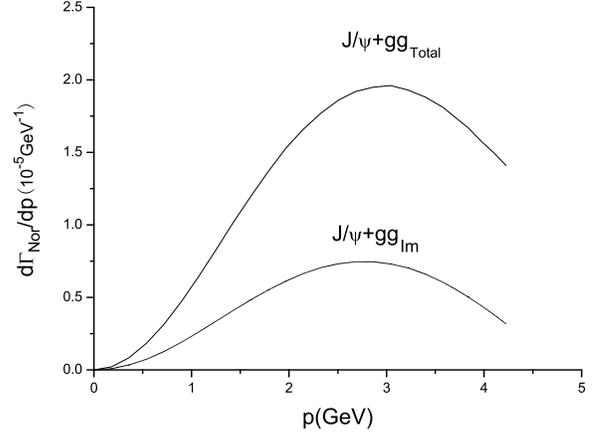}
\caption{The normalized partial width of the $\mathcal{O}(\alpha_s^6)$ $\Upsilon\to J/\psi+gg$ process as
function of $J/\psi$ momentum $p_{J/\psi}$. The solid line is the total result, and the dashed line
is the contribution from the imaginary  part in the Feynman amplitude.}
\end{center}
\end{figure}

The experimental result in Eq.(\ref{exp}) includes the feed-down contributions of $\chi_{cJ}$, which are
$<8.2,11,10$ percents for $J=0,1,2$ respectively, and $24\%$ feed-down contribution of $\psi(2S)$.
Removing the feed-down contributions, the branching ratio of direct $J/\psi$ production in $\Upsilon$
decay would be
\begin{equation}
\mathcal{B}(\Upsilon\to J/\psi_{\mathrm{direct}}+X)=3.52\times10^{-4}.
\end{equation}
Which is about 3.8 times larger than the current CS prediction.

For the $J/\psi$ momentum spectrum, the
normalized decay widths defined in Eq.(\ref{normal}) are used to present the results with
default parameter choice. Both the total result (solid line) and the
imaginary part contribution (dashed line) for $\mathcal{O}(\alpha_s^6)$ $J/\psi+gg$ process
are shown in Fig.[3] with very similar shape.
A summarized CS contribution to the $p_{J/\psi}$ distribution of the normalized decay
width is shown in Fig.[4]. And we find the peak of total result curve is at $p_{J/\psi}=2.7\mathrm{GeV}$,
which is a little larger than that of the CLEO measurement\cite{Briere:2004ug}.
\begin{figure}
\begin{center}
\includegraphics[scale=0.80]{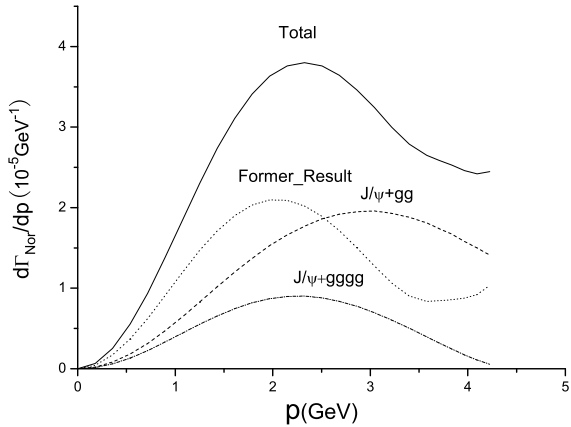}
\caption{The normalized partial width for the CS $J/\psi$ production in $\Upsilon$ decay as
function of $J/\psi$ momentum $p_{J/\psi}$. The solid line is the total result. the dashed line
is the contribution of the $\mathcal{O}(\alpha_s^6)$ $\Upsilon\to J/\psi+gg$ process.
The dashed-dot line is the 100$\times$ the contribution $\mathcal{O}(\alpha_s^6)$ $\Upsilon\to J/\psi+gggg$.
The dot line is the contribution calculated in Ref.\cite{He:2009by}, which includes the $\mathcal{O}(\alpha_s^5)$
$\Upsilon\to J/\psi+c\bar{c}g$ and $\mathcal{O}(\alpha^2\alpha_s^2)$ $\Upsilon\to J/\psi+gg$ and
$\Upsilon\to J/\psi+c\bar{c}$ processes.}
\end{center}
\end{figure}

It is found that the $J/\psi$ production in association with $c\bar{c}$ pair is an important mechanism
for $J/\psi$ electroproduction\cite{Abe:2002rb,:2009nj} in the Belle experiment. And theoretically, the contribution of
$p\bar{p}\to J/\psi+c\bar{c}+X$ process to $J/\psi$ hadroproduction at the Tevatron is also found to be
non-ignorable\cite{Artoisenet:2007xi,He:2009zzb}. The ratio of $J/\psi$ production in association with $c\bar{c}$ pair to
$J/\psi$ plus anything may also be a good probe to reveal the $J/\psi$ production mechanism in $\Upsilon$ decay
and to clarify the conflict between the CLEO measurement and theoretical prediction.
Choosing $\alpha_s(2m_c)=0.259$, we give the CS prediction for the ratio $R_{cc}$
\begin{equation}\label{ratioc}
R_{cc}=\frac{\mathcal{B}(\Upsilon\to J/\psi+c\bar{c}+X)}{\mathcal{B}(\Upsilon\to J/\psi+X)}=0.39^{+0.21}_{-0.20},
\end{equation}
where the center, upper and lower bound values correspond to $r=0.327,0.296$ and 0.361 respectively,
and the associated charmed particles process includes the $\mathcal{O}(\alpha_s^5)$
$\Upsilon\to J/\psi+c\bar{c}g$ sub-process, which is dominant, and $\mathcal{O}(\alpha^2\alpha_s^2)$
$\Upsilon\to \gamma^{\ast}\to J/\psi+c\bar{c}$ sub-process. On the contrary, the CO prediction of
$R_{cc}$ is only at the level of $1\%$\cite{Cheung:1996mh}, which is quite different with the CS prediction.
Unlike the branching ratio, the theoretical prediction of $R_{cc}$ only depends on $r$ and $\alpha_s$,
which results in a relatively small uncertainty. Particularly, if we drop the contribution of QED part,
$R_{cc}$ is just proportional to $\alpha_s$. In Ref.\cite{Zhang:2005cha,Gong:2009ng}, the authors find
the enhancement of the NLO QCD corrections is large in $e^{+}e^{-}\to J/\psi+c\bar{c}$ process.
It indicates that the result in Eq.[\ref{ratioc}] is only a very preliminary result and to get a more solid
predictions the contribution of the NLO QCD corrections to $\Upsilon\to J/\psi+c\bar{c}+g$ process should
be taken into account. Calculating the NLO QCD corrections to $\Upsilon\to J/\psi+c\bar{c}+g$ process is
beyond the scope of this work and will not be discussed here. In the end, the $J/\psi$ momentum spectra for
the associated process and $\mathrm{non}-c\bar{c}$ process are given in Fig.[5] for comparison.

\begin{figure}
\begin{center}
\includegraphics[scale=0.80]{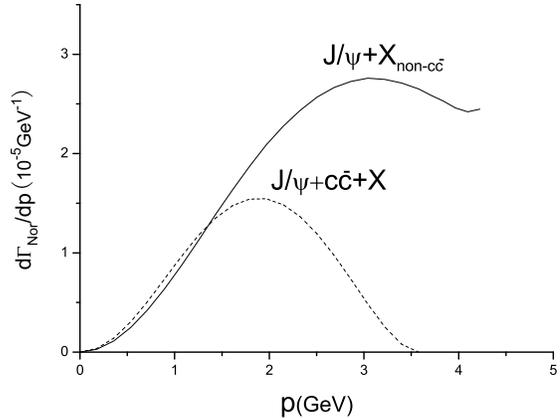}
\caption{The normalized partial width of $\Upsilon$ decay into $J/\psi+c\bar{c}+X$ process  as
function of $J/\psi$ momentum $p_{J/\psi}$ (dashed line) and that for $J/\psi+X$ production (solid line).}
\end{center}
\end{figure}

In summary, in this work, we calculate the $\mathcal{O}(\alpha_s^6)$ CS contribution of $\Upsilon\to J/\psi+gg$
and $\Upsilon\to J/\psi+gggg$ processes to the inclusive $J/\psi$ production in $\Upsilon$ decay. The branching
ratio is estimated in two ways. In the first way, the numerical results of partial width and branching
ratios are all evaluated directly. And we find its result is more close to the experimental data when the values
of the parameters are properly chosen, which is also coincident with a rough estimated result mentioned in Ref.\cite{Trottier:1993ze}.
However, the uncertainty of this way is very large and the branching can be in a wide range of $2.3\times10^{-5}\sim 1.7\times10^{-4}$.
In the second way, the branching ratio is calculated by using the normalized decay width, which seems more reliable.
After combining the present result  with the contribution calculated in our previous work\cite{He:2009by}, we
find now the total CS prediction is about $0.92\times10^{-4}$, which is still about 3.8 times less than the experimental
value $3.2\times10^{-4}$ for direct $J/\psi$ production given by the CLEO Collaboration\cite{Briere:2004ug}.
We also calculate the $J/\psi$ momentum spectrum and find the peak of the color-singlet curves is close to that of the
CLEO result, although being a little larger. Besides the the branching ratio and the $J/\psi$ spectrum, we also study
the ratio $R_{cc}=\mathcal{B}(\Upsilon\to J/\psi+c\bar{c}+X)/\mathcal{B}(\Upsilon\to J/\psi+X)$, and find the CS prediction
is much larger than that of the CO. Since the associated charmed meson process can be measured separately, a analysis of the ratio is expected to perform in the CLEO, Babar or Belle experiments.
Now there is still large discrepancy between the CLEO results and NRQCD predictions. There are two points to be addressed: first the $J/\psi$ production mechanism is
not well understood yet, and the existence of the CO mechanism is still under debate; second the higher QCD
corrections are not included completely. Therefore, to understand the $J/\psi$ production mechanism in $\Upsilon$ decay
and moreover in $p\bar{p}$ collisions at the Tevatron. Further theoretical and experimental work are necessary.

\section*{Acknowledgement}
This work is supported by the National Natural Science Foundation of
China (No. 10979056 and 10935012), and by the Chinese Academy of Science under
Project No. INFO-115-B01.
The work of Zhiguo He is partially supported by the CPAN08-PD14
contract of the CSD2007-00042 Consolider-Ingenio 2010 program, and
by the FPA2007-66665-C02-01/ project (Spain).


\begin{thebibliography}{99}
%\cite{Bodwin:1994jh}
\bibitem{Bodwin:1994jh}
  G.~T.~Bodwin, E.~Braaten and G.~P.~Lepage,
  %``Rigorous QCD analysis of inclusive annihilation and production of heavy
  %quarkonium,''
  Phys.\ Rev.\  D {\bf 51}, 1125 (1995)
  [Erratum-ibid.\  D {\bf 55}, 5853 (1997)].
%  [arXiv:hep-ph/9407339].
  %%CITATION = PHRVA,D51,1125;%%


%\cite{Abe:2002rb}
\bibitem{Abe:2002rb}
  K.~Abe {\it et al.}  [Belle Collaboration],
  %``Observation of double c anti-c production in e+ e- annihilation at
  %s**(1/2) approx. 10.6-GeV,''
  Phys.\ Rev.\ Lett.\  {\bf 89}, 142001 (2002).
%  [arXiv:hep-ex/0205104].
  %%CITATION = PRLTA,89,142001;%%


%\cite{:2009nj}
\bibitem{:2009nj}
  P.~Pakhlov {\it et al.}  [Belle Collaboration],
  %``Measurement of the e+e- -> J/psi ccbar cross section at \sqrt{s} ~10.6
  %GeV,''
  Phys.\ Rev.\  D {\bf 79}, 071101 (2009).
%  [arXiv:0901.2775 [hep-ex]].
  %%CITATION = PHRVA,D79,071101;%%


%\cite{Zhang:2005cha}
\bibitem{Zhang:2005cha}
  Y.~J.~Zhang, Y.~j.~Gao and K.~T.~Chao,
  %``Next-to-leading order QCD correction to e+ e- --> J/psi + eta/c at
  %s**(1/2) = 10.6-GeV,''
  Phys.\ Rev.\ Lett.\  {\bf 96}, 092001 (2006).
%  [arXiv:hep-ph/0506076].
  %%CITATION = PRLTA,96,092001;%%


%\cite{Gong:2009ng}
\bibitem{Gong:2009ng}
  B.~Gong and J.~X.~Wang,
  %``Next-to-leading-order QCD corrections to $e^+e^- \to J/\psi_{cc} $  at the
  %$B$ factories,''
  Phys.\ Rev.\  D {\bf 80}, 054015 (2009).
%  [arXiv:0904.1103 [hep-ph]].
  %%CITATION = PHRVA,D80,054015;%%


%\cite{Ma:2008gq}
\bibitem{Ma:2008gq}
  Y.~Q.~Ma, Y.~J.~Zhang and K.~T.~Chao,
  %``QCD correction to $\bm{e^+ e^- \to J/\psi g g}$ at B Factories,''
  Phys.\ Rev.\ Lett.\  {\bf 102}, 162002 (2009).
%  [arXiv:0812.5106 [hep-ph]].
  %%CITATION = PRLTA,102,162002;%%


%\cite{Gong:2009kp}
\bibitem{Gong:2009kp}
  B.~Gong and J.~X.~Wang,
  %``Next-to-Leading-Order QCD Corrections to e^ + e^- -> J/\psi+gg at the B
  %Factories,''
  Phys.\ Rev.\ Lett.\  {\bf 102}, 162003 (2009).
%  [arXiv:0901.0117 [hep-ph]].
  %%CITATION = PRLTA,102,162003;%%

%\cite{He:2007te}
\bibitem{He:2007te}
  Z.~G.~He, Y.~Fan and K.~T.~Chao,
  %``Relativistic corrections to $J/\psi$ exclusive and inclusive double   charm
  %production at B factories,''
  Phys.\ Rev.\  D {\bf 75}, 074011 (2007).
%  [arXiv:hep-ph/0702239].
  %%CITATION = PHRVA,D75,074011;%%

%\cite{He:2009uf}
\bibitem{He:2009uf}
  Z.~G.~He, Y.~Fan and K.~T.~Chao,
  %``Relativistic correction to $e^{+}e^{-}\to J/\psi+gg$ at $B$ factories and
  %constraint on color-octet matrix elements,''
  Phys.\ Rev.\  D {\bf 81}, 054036 (2010)
%  [arXiv:0910.3636 [hep-ph]].
  %%CITATION = PHRVA,D81,054036;%%
%\cite{Jia:2009np}
%\bibitem{Jia:2009np}
  Y.~Jia,
  %``Color-singlet relativistic correction to inclusive $J/\psi$ production
  %associated with light hadrons at $B$ factories,''
  Phys.\ Rev.\  D {\bf 82}, 034017 (2010)
%  [arXiv:0912.5498 [hep-ph]].
  %%CITATION = PHRVA,D82,034017;%%

%\cite{Zhang:2009ym}
\bibitem{Zhang:2009ym}
  Y.~J.~Zhang, Y.~Q.~Ma, K.~Wang and K.~T.~Chao,
  %``QCD radiative correction to color-octet $J/\psi$ inclusive production at B
  %Factories,''
  arXiv:0911.2166 [hep-ph].
  %%CITATION = ARXIV:0911.2166;%%

%\cite{Li:2010xu}
\bibitem{Li:2010xu}
  R.~Li and J.~X.~Wang,
  %``The next-to-leading-order QCD correction to inclusive J/\psi(\Upsilon)
  %production in Z^0 decay,''
  arXiv:1007.2368 [hep-ph].
  %%CITATION = ARXIV:1007.2368;%%

%\cite{Artoisenet:2009xh}
\bibitem{Artoisenet:2009xh}
  P.~Artoisenet, J.~M.~Campbell, F.~Maltoni and F.~Tramontano,
  %``J/psi production at HERA,''
  Phys.\ Rev.\ Lett.\  {\bf 102}, 142001 (2009)
  [arXiv:0901.4352 [hep-ph]].
  %%CITATION = PRLTA,102,142001;%%


%\cite{Chang:2009uj}
\bibitem{Chang:2009uj}
  C.~H.~Chang, R.~Li and J.~X.~Wang,
  %``J/\psi polarization in photo-production up-to the next-to-leading order of
  %QCD,''
  Phys.\ Rev.\  D {\bf 80}, 034020 (2009)
  [arXiv:0901.4749 [hep-ph]].
  %%CITATION = PHRVA,D80,034020;%%

%\cite{Butenschoen:2009zy}
\bibitem{Butenschoen:2009zy}
  M.~Butenschoen and B.~A.~Kniehl,
  %``Complete next-to-leading-order corrections to J/psi photoproduction in
  %nonrelativistic quantum chromodynamics,''
  Phys.\ Rev.\ Lett.\  {\bf 104} (2010) 072001
  [arXiv:0909.2798 [hep-ph]].
  %%CITATION = PRLTA,104,072001;%%

%\cite{Butenschoen:2009zy}
\bibitem{Butenschoen:2009zy}
  M.~Butenschoen and B.~A.~Kniehl,
  %``Complete next-to-leading-order corrections to J/psi photoproduction in
  %nonrelativistic quantum chromodynamics,''
  arXiv:0909.2798 [hep-ph].
  %%CITATION = ARXIV:0909.2798;%%


%\cite{Campbell:2007ws}
\bibitem{Campbell:2007ws}
  J.~M.~Campbell, F.~Maltoni and F.~Tramontano,
  %``QCD corrections to J/psi and Upsilon production at hadron colliders,''
  Phys.\ Rev.\ Lett.\  {\bf 98}, 252002 (2007)
  [arXiv:hep-ph/0703113].
  %%CITATION = PRLTA,98,252002;%%


%\cite{Gong:2008sn}
\bibitem{Gong:2008sn}
  B.~Gong and J.~X.~Wang,
  %``Next-to-leading-order QCD corrections to $J/\psi$ polarization at Tevatron
  %and Large-Hadron-Collider energies,''
  Phys.\ Rev.\ Lett.\  {\bf 100}, 232001 (2008)
  [arXiv:0802.3727 [hep-ph]].
  %%CITATION = PRLTA,100,232001;%%


%\cite{Gong:2008hk}
\bibitem{Gong:2008hk}
  B.~Gong and J.~X.~Wang,
  %``QCD corrections to polarization of $J/\psi$ and $\upsilon$ at Tevatron and
  %LHC,''
  Phys.\ Rev.\  D {\bf 78}, 074011 (2008)
  [arXiv:0805.2469 [hep-ph]].
  %%CITATION = PHRVA,D78,074011;%%


%\cite{Braaten:1994vv}
\bibitem{Braaten:1994vv}
  E.~Braaten and S.~Fleming,
  %``Color octet fragmentation and the psi-prime surplus at the Tevatron,''
  Phys.\ Rev.\ Lett.\  {\bf 74}, 3327 (1995)
  [arXiv:hep-ph/9411365].
  %%CITATION = PRLTA,74,3327;%%


%\cite{Gong:2008ft}
\bibitem{Gong:2008ft}
  B.~Gong, X.~Q.~Li and J.~X.~Wang,
  %``QCD corrections to $J/\psi$ production via color octet states at Tevatron
  %and LHC,''
  Phys.\ Lett.\  B {\bf 673}, 197 (2009)
  [arXiv:0805.4751 [hep-ph]].
  %%CITATION = PHLTA,B673,197;%%


%\cite{Cheung:1996mh}
\bibitem{Cheung:1996mh}
  K.~m.~Cheung, W.~Y.~Keung and T.~C.~Yuan,
  %``Color octet $J/\psi$ production in the $\upsilon$ decay,''
  Phys.\ Rev.\  D {\bf 54}, 929 (1996)
  [arXiv:hep-ph/9602423].
  %%CITATION = PHRVA,D54,929;%%



%\cite{Napsuciale:1997bz}
\bibitem{Napsuciale:1997bz}
  M.~Napsuciale,
  %``Inclusive $J/\psi$ production in $\upsilon$ decay via color octet
  %mechanisms,''
  Phys.\ Rev.\  D {\bf 57}, 5711 (1998)
  [arXiv:hep-ph/9710488].
  %%CITATION = PHRVA,D57,5711;%%


%\cite{Liu:2009ra}
\bibitem{Liu:2009ra}
  X.~Liu,
  %``Color-Octet $\jpsi$ Production in $\Upsilon$ Decay near the Kinematic
  %Limit,''
  arXiv:0909.2565 [hep-ph].
  %%CITATION = ARXIV:0909.2565;%%


%\cite{Trottier:1993ze}
\bibitem{Trottier:1993ze}
  H.~D.~Trottier,
  %``Upsilon decay into charmonium and the color octet mechanism,''
  Phys.\ Lett.\  B {\bf 320}, 145 (1994)
  [arXiv:hep-ph/9307315].
  %%CITATION = PHLTA,B320,145;%%



%\cite{Li:1999ar}
\bibitem{Li:1999ar}
  S.~y.~Li, Q.~b.~Xie and Q.~Wang,
  %``Contribution of colour-singlet process Upsilon --> J/psi + c anti-c g  to
  %Upsilon --> J/psi + X,''
  Phys.\ Lett.\  B {\bf 482}, 65 (2000)
  [arXiv:hep-ph/9912328].
  %%CITATION = PHLTA,B482,65;%%

%\cite{Han:2006vi}
\bibitem{Han:2006vi}
  W.~Han and S.~Y.~Li,
  %``Comments on CLEO new measurements for $\upsilon_{1S}$ decays to charmonium
  %final states and investigations on associate strange particle enhancement in
  %$\upsilon$ to J/Psi + $X$,''
  Phys.\ Rev.\  D {\bf 74}, 117502 (2006)
  [arXiv:hep-ph/0607251].
  %%CITATION = PHRVA,D74,117502;%%



%\cite{He:2009by}
\bibitem{He:2009by}
  Z.~G.~He and J.~X.~Wang,
  %``Inclusive J/\psi Production In \Upsilon Decay Via Color-singlet
  %Mechanism,''
  arXiv:0911.0139 [hep-ph].
  %%CITATION = ARXIV:0911.0139;%%

%\cite{Fulton:1988ug}
\bibitem{Fulton:1988ug}
  R.~Fulton {\it et al.}  [CLEO Collaboration],
  %``FIRST OBSERVATION OF INCLUSIVE PSI PRODUCTION IN UPSILON DECAYS,''
  Phys.\ Lett.\  B {\bf 224}, 445 (1989).
  %%CITATION = PHLTA,B224,445;%%


%\cite{Maschmann:1989ai}
\bibitem{Maschmann:1989ai}
  W.~S.~Maschmann {\it et al.}  [Crystal Ball Collaboration],
  %``Inclusive J / Psi Production In Decays Of B Mesons,''
  Z.\ Phys.\  C {\bf 46}, 555 (1990).
  %%CITATION = ZEPYA,C46,555;%%


%\cite{Albrecht:1992ap}
\bibitem{Albrecht:1992ap}
  H.~Albrecht {\it et al.}  [ARGUS Collaboration],
  %``Search For Charm Production In Direct Decays Of The Upsilon (1s)
  %Resonance,''
  Z.\ Phys.\  C {\bf 55}, 25 (1992).
  %%CITATION = ZEPYA,C55,25;%%


%\cite{Briere:2004ug}
\bibitem{Briere:2004ug}
  R.~A.~Briere {\it et al.}  [CLEO Collaboration],
  %``New measurements of $\upsilon_{1S}$ decays to charmonium final states,''
  Phys.\ Rev.\  D {\bf 70}, 072001 (2004)
  [arXiv:hep-ex/0407030].
  %%CITATION = PHRVA,D70,072001;%%

\bibitem{FDC}
  J.-X.~Wang,
  Nucl.\ Instrum.\ Meth.\ A {\bf 534}, 241 (2004).


%\cite{Amsler:2008zzb}
\bibitem{Amsler:2008zzb}
  C.~Amsler {\it et al.}  [Particle Data Group],
  %``Review of particle physics,''
  Phys.\ Lett.\  B {\bf 667}, 1 (2008).
  %%CITATION = PHLTA,B667,1;%%

%\cite{Mackenzie:1981sf}
\bibitem{Mackenzie:1981sf}
  P.~B.~Mackenzie and G.~P.~Lepage,
  %``QCD Corrections To The Gluonic Width Of The Upsilon Meson,''
  Phys.\ Rev.\ Lett.\  {\bf 47}, 1244 (1981).
  %%CITATION = PRLTA,47,1244;%%


%\cite{Artoisenet:2007xi}
\bibitem{Artoisenet:2007xi}
  P.~Artoisenet, J.~P.~Lansberg and F.~Maltoni,
  %``Hadroproduction of $J/\psi$ and $\upsilon$ in association with a
  %heavy-quark pair,''
  Phys.\ Lett.\  B {\bf 653}, 60 (2007)
  [arXiv:hep-ph/0703129].
  %%CITATION = PHLTA,B653,60;%%

%\cite{He:2009zzb}
\bibitem{He:2009zzb}
  Z.~G.~He, R.~Li and J.~X.~Wang,
  %``QED contribution to the production of J/\psi+c\bar{c}+X at the Tevatron and
  %LHC,''
  Phys.\ Rev.\  D {\bf 79}, 094003 (2009)
  [arXiv:0904.2069 [hep-ph]].
  %%CITATION = PHRVA,D79,094003;%%


\end{thebibliography}
\end{document}